\documentclass[12pt]{iopart}
\usepackage{iopams}  
\usepackage{setstack}
\usepackage{amssymb}
\usepackage{amsfonts}

\newtheorem{lemma}{Lemma}
\newtheorem{theorem}{Theorem}
\newtheorem{definition}{Definition}

\begin{document}
\title[Algebraic Properties of Quasilinear Two-Dimensional Lattices ]{Algebraic Properties of Quasilinear Two-Dimensional Lattices  connected with integrability}

\author{I.T. Habibullin$^{1,2}$, M.N. Kuznetsova$^1$}

\address{$^1$Institute of Mathematics, Ufa Federal Research Centre, Russian Academy of Sciences,
112, Chernyshevsky Street, Ufa 450008, Russian Federation}
\address{$^2$Bashkir State University, 32 Validy Street, Ufa 450076 , Russian Federation} 
\eads{\mailto{habibullinismagil@gmail.com}, \mailto{mariya.n.kuznetsova@gmail.com}}

\begin{abstract}

In the article a classification method for nonlinear integrable equations with three independent variables is discussed based on the notion of the integrable reductions. We call the equation integrable if it admits a large class of reductions being Darboux integrable systems of hyperbolic type equations with two independent variables. The most natural and convenient object to be studied within the frame of this scheme is the class of two dimensional lattices generalizing the well-known Toda lattice. In the present article we deal with the quasilinear lattices of the form $u_{n,xy}=\alpha(u_{n+1} ,u_n,u_{n-1} )u_{n,x}u_{n,y} + \beta(u_{n+1},u_n,u_{n-1})u_{n,x}+\gamma(u_{n+1} ,u_n,u_{n-1} )u_{n,y}+\delta(u_{n+1} ,u_n,u_{n-1})$. We specify the coefficients of the lattice assuming that there exist cutting off conditions which reduce the lattice to a Darboux integrable hyperbolic type system of the arbitrarily high order. Under some extra assumption of nondegeneracy we described the class of the lattices integrable in the sense indicated above. There are new examples in the obtained list of chains.

\end{abstract}

\maketitle

\eqnobysec

\section{Introduction}

Integrable equations with three independent variables have a wide range of applications in physics. It suffices to recall such well-known nonlinear models as the KP equation, the Davey-Stewartson  equation, the Toda lattice equation, and so on. From the point of view of integration and classification, multidimensional equations are the most complex. Different approaches to  study the integrable multidimensional models are discussed, for example, in the papers \cite{Ferapontov2004}--\cite{Sergeev}.
It is known that the symmetry approach \cite{Adler,Mikhailov91}, which has proved to be a very effective method for classifying integrable equations in 1 + 1 dimensions, is not so effective in the multidimensionality \cite{Mikhailov98}. For studying multidimensional equations, the idea of the reduction is often used, when the researches replace the equation with a system of equations with fewer independent variables. The existence of a wide class of integrable reductions with two independent variables, as a rule, indicates the integrability of an equation with three independent variables. Among the specialists, the most popular method is the method of hydrodynamic reductions, when the presence of an infinite set of integrable systems of hydrodynamic type is taken as a sign of integrability of the equation, the general solution of each of which generates some solution of the equation under consideration (see, for example, \cite{FKhP,Ferapontov2004,Ferapontov2006}). 
The history of the method and related references can be found in the survey \cite {Sokolov}.

In our works \cite{H2013,HabPoptsovaSIGMA17} we use an alternative approach. We call this equation integrable if it admits an infinite class of reductions in the form of Darboux-integrable systems of partial differential equations of hyperbolic type with two independent variables.
In solving classification problems for multidimensional equations, the apparatus of characteristic Lie algebras can be used in this formulation (a detailed exposition can be found in \cite{ZMHSbook,Sh1995}). This direction in the theory of integrability seems to us promising.
Consider a nonlinear chain 
\begin{equation} \label{eq0}  
u_{n,xy}=f(u_{n+1},u_n,u_{n-1}, u_{n,x},u_{n,y})
\end{equation}
with three independent variables, where the sought function $ u = u_ {n} (x, y) $ depends on the real $ x $, $ y $, and the integer $ n $. For the chain (\ref{eq0}), the desired finite-field reductions are obtained in a natural way, in a sufficiently suitable way to break off the chain at two integer points
\begin{eqnarray}
&&u_{N_1} =\varphi_1(x,y,u_{N_1+1},...), \label{bc1}\\
&&u_{n,xy}=f(u_{n+1},u_n,u_{n-1}, u_{n,x},u_{n,y}),\quad N_1 < n < N_2, \label{eq00} \\
&&u_{N_2} =\varphi_2(x,y,u_{N_2-1},...). \label{bc2}
\end{eqnarray}   
Examples of such boundary conditions can be found below (see (\ref{bc11}), (\ref{bc21})).
The following two very significant circumstances should be noted: 
\begin{itemize}
\item[i)] for any known integrable chain of the form (\ref{eq0}) there are cut-off conditions reducing it to a Darboux-integrable system of the form (\ref{bc1})-(\ref{bc2}) of arbitrarily large order $ N = N_2-N_1-1 $; 
\item[ii)] specific form of the functions $ \varphi_1 $, $ \varphi_2 $, and $ f $ is constructively determined by the requirement of integrability of the system in the sense of Darboux.
\end{itemize}

These two facts serve as motivation for the following definition (see also the work \cite {H2013}):
\begin{definition}
A chain (\ref{eq0}) is called integrable if there exist functions $\varphi_1$ and $\varphi_2$ such that for any choice of a pair of integers $ N_1 $, $ N_2 $, where $ N_1 <N_2-1 $, the hyperbolic type system 
(\ref{bc1})-(\ref{bc2}) is Darboux integrable.
\end{definition}

In the present paper we investigate quasilinear chains of the following form
\begin{equation} \label{eq1} 
u_{n,xy}=\alpha u_{n,x}u_{n,y} + \beta u_{n,x}+\gamma u_{n,y}+\delta,
\end{equation}
assuming that the functions $\alpha=\alpha(u_{n+1} ,u_n,u_{n-1} )$, $\beta=\beta(u_{n+1} ,u_n,u_{n-1} )$,\\ $\gamma=\gamma(u_{n+1} ,u_n,u_{n-1} )$, $\delta=\delta(u_{n+1} ,u_n,u_{n-1} )$ are analytic in the domain $D\subset \mathbb{C}^3$. We also assume that the derivatives  
\begin{equation} \label{eq11} 
 \frac{\partial\alpha(u_{n+1} ,u_n,u_{n-1} )}{\partial u_{n+1} }\quad \mbox{我} \quad\frac{\partial\alpha(u_{n+1} ,u_n,u_{n-1} )}{\partial u_{n-1} } 
\end{equation}
differ from zero.

The main result of this paper is the proof of the following assertion
\begin{theorem}\label{th0}
The quasilinear chain (\ref{eq1}), (\ref{eq11}) is integrable in the sense of Definition 1 if and only if it is reduced by point transformations to one of the following forms 
\begin{eqnarray} 
&i)&u_{n,xy}=\alpha_nu_{n,x}u_{n,y},\nonumber\\
&ii)&u_{n,xy}=\alpha_n(u_{n,x}u_{n,y}-u_n(u_{n,x}+u_{n,y})+u_n^2) + u_{n,x}+u_{n,y}-u_n ,\nonumber \\
&iii)&u_{n,xy}=\alpha_n(u_{n,x}u_{n,y}-s_{n}(u_{n,x}+u_{n,y})+s_{n}^2) + s'_{n}(u_{n,x}+u_{n,y}-s_n),\nonumber
\end{eqnarray}
where
$$ s_{n}=u_n^2+C,\quad s'_n=2u_n,\quad \alpha_n = \frac{1}{u_n - u_{n-1}} - \frac{1}{u_{n+1}-u_n},$$
$C$ is an arbitrary constant.
\end{theorem}
We note that equation i) was found earlier in the papers \cite {Fer-TMF}, \cite {ShY} by Ferapontov and Shabat and Yamilov, equations ii) and iii) appear to be new. By applying additional conditions of the form $ x = \pm y $ to the equations i)-iii), we obtain 1 + 1 -dimensional integrable chains. It is easily verified that by point transformations they are reduced to the equations found earlier  by Yamilov (see \cite {YamilovJPA06}).

Following Definition 1, we suppose that there are cut-off conditions such that by imposing them at two arbitrary points $ n = N_1 $, $ n = N_2 $ ($ N_1 <N_2-1 $) to  the chain (\ref{eq1}) we obtain a system of hyperbolic type equations
\begin{eqnarray}
&&u_{N_1} =\varphi_1, \nonumber\\
&&u_{n,xy}= \alpha_n u_{n,x}u_{n,y}+\beta_n u_{n,x} + \gamma_n u_{n,y} + \delta_n,\quad N_1 < n < N_2, \label{eq3} \\
&&u_{N_2} =\varphi_2. \nonumber
\end{eqnarray}
that is integrable in the sense of Darboux. 

We recall that the system of partial differential equations of the hyperbolic type (\ref{eq3}) is Darboux integrable if it has a complete set of functionally independent $ x- $ and $ y- $ integrals. A function $ I $ that depends on a finite set of dynamical variables $ {\bf{u}}, {\bf {u}}_x, {\bf {u}}_ y, \ldots $ is called a $ y $ -integral if it satisfies the equation $ D_y I = 0 $, where $ D_y $ is the operator of total derivative with respect to the variable $ y $, and the vector $ \bf {u} $ has the coordinates $ u_ {N_1 + 1}, u_ {N_1 + 2}, \ldots, u_{N_2-1} $. Since the system (\ref{eq3}) is autonomous, we consider only autonomous nontrivial integrals. It can be shown that the $ y- $ integral does not depend on $ {\bf u} _y, {\bf u} _ {yy}, \ldots $. Therefore, we will consider only $ y- $ integrals that depend on at least one dynamic variable $ {\bf u}, {\bf u} _x, \ldots $. We note that nowadays the Darboux integrable discrete and continuous models are intensively studied (see, \cite {H2013, ZMHSbook}, \cite{H2007}-\cite{Yamilov}).

We give one more argument in favor of our Definition 1 concerning the integrability property of a two-dimensional chain. The problem of finding a general solution of a Darboux-integrable system reduces to the problem of solving a system of ordinary differential equations. Usually these ODEs are solved explicitly. On the other hand, any solution of the considered hyperbolic system (\ref{eq3}) easily extends beyond the interval $ [N_1, N_2] $ and generates the solution of the corresponding chain (\ref{eq1}). Therefore, in this case the chain (\ref{eq1}) has a large set of exact solutions.

Let us briefly explain the structure of the paper. In \S 2 we recall the necessary definitions and investigate the basic properties of the characteristic Lie algebra, which is the main tool in the theory of Darboux-integrable systems.
In \S 3 we introduce the definition of test sequences, by means of which we obtain a system of differential equations for the refinement of the functions $ \alpha $, $ \beta $, $ \gamma $. Paragraph 4 is devoted to the search for the function $ \delta $. Here we also give the final form of the desired chain (\ref{res_latt}) that is integrable in the sense of Definition 1 and the proof of the Theorem~\ref{th0} is given.

\section{Characteristic Lie algebras}

Since the chain (\ref{eq1}) is invariant under the shift of the variable $ n $, then without loss of generality we can put $ N_1 = -1 $. In what follows we consider a system of hyperbolic equations
\begin{eqnarray}
&&u_{-1}  = \varphi_1, \nonumber\\
&&u_{n,xy} =\alpha_n u_{n,x}u_{n,y}+\beta_n u_{n,x} + \gamma_n u_{n,y} + \delta_n, \quad 0 \leq n \leq N, \label{eq2_1} \\
&&u_{N+1}  = \varphi_2. \nonumber
\end{eqnarray}
Recall that here $\alpha_n = \alpha(u_{n-1},u_n,u_{n+1})$,  $\beta_n = \beta(u_{n-1},u_n,u_{n+1})$,  $\gamma_n = \gamma(u_{n-1},u_n,u_{n+1})$,  $\delta_n = \delta(u_{n-1},u_n,u_{n+1})$. Suppose that the system (\ref{eq2_1}) is Darboux integrable and that $ I ({\bf u}, {\bf u} _x, \ldots) $ is its nontrivial $ y $ -integral. The latter means that the function $ I $ must satisfy the equation $ D_y I = 0 $, where $ D_y $ is the operator of total derivative  with respect to the variable $ y $. The operator $ D_y $ acts on the class of functions of the form $ I ({\bf u}, {\bf u} _x, \ldots) $ due to  the rule $ D_y I = Y I $, where

\begin{equation} \label{eq2_2} 
Y = \sum_{i=0} ^N \left(u_{i,y} \frac{\partial}{\partial u_i} + f_i \frac{\partial}{\partial u_{i,x}} + f_{i,x}\frac{\partial}{\partial u_{i,xx}} + \cdots  \right).
\end{equation} 
Here $f_i = \alpha_i u_{i,x} u_{i,y} + \beta_i u_{i,x} + \gamma_i u_{i,y} + \delta_i$ is the right hand side of the lattice (\ref{eq1}). Hence, the function $ I $ satisfies the equation $ Y I = 0 $. The coefficients of the equation $ YI = 0 $ depend on the variables $ u_ {i, y} $, while its solution $ I $ does not depend on $ u_ {i, y} $, therefore the function $ I $ actually satisfies the system linear equations:
\begin{equation} \label{eq2_3} 
YI=0, \quad X_j I = 0, \quad j=1,\ldots,N,
\end{equation}
where $X_i = \frac{\partial}{\partial u_{i,y}}$. It follows from (\ref{eq2_3}) that the commutator $Y_i=[X_i,Y]$ of the operators $ Y $ and $ X_i $ for $ i = 0,1, ... N $ also annuls $ I $. We use the explicit coordinate representation of the operator $ Y_i $:
\begin{equation} \label{eq2_4} 
Y_i = \frac{\partial}{\partial u_i} + X_i(f_i) \frac{\partial}{\partial u_{i,x}} + X_i(D_x f_i) \frac{\partial}{\partial u_{i,xx}} + \cdots
\end{equation} 
By the special form of the function $ f_i $, the operator $ Y $ can be represented in the form:
\begin{equation} \label{eq2_5}
Y=\sum_{i=0}^N u_{i,y} Y_i + R,
\end{equation}
where
\begin{eqnarray} \label{op_R}
&&R=\sum_{i=0}^N (f_i - u_{i,y}X_i(f_i))\frac{\partial}{\partial u_{i,x}} + (f_{i,x}-u_{i,y}X_i(D_xf_i))\frac{\partial}{\partial u_{i,xx}}+\cdots=\nonumber\\
&&\hphantom{R}= \sum_{i=0}^N (\beta_i u_{i,x}+\delta_i)\frac{\partial}{\partial u_{i,x}} + \nonumber\\
&&\hphantom{R=\sum_{i=0}^N}+\bigl((\alpha_i u_{i,x}+\gamma_i)(\beta_i u_{i,x}+\delta_i)+D_x(\beta_i u_{i,x}+\delta_i)\bigr)\frac{\partial}{\partial u_{i,xx}}+\cdots
\end{eqnarray}

Denote by $ \bf{F} $ the ring of locally analytic functions of the dynamical variables $ {\bf{u}}, {\bf{u}}_x, {\bf{u}}_y, \ldots $. 

Consider the Lie algebra $ \mathcal {L} (y, N) $ over the ring $ \bf{F} $ generated by the differential operators $ Y, Y_0, Y_1, ..., Y_N $. It is clear that the operations of computing the commutator of two vector fields and multiplying the vector field by a function satisfy the following conditions:
\begin{eqnarray}
&&[Z,gW]=Z(g)W+g[Z,W], \label{rh1}\\
&&(gZ)h=gZ(h),\label{rh2}
\end{eqnarray}
where $Z,W\in\mathcal{L}(y,N)$, $g,h\in\bf{F}$. Consequently, the pair $ (\bf{F}, \mathcal{L} (y, N)) $ has the structure of the Lie-Rinehart algebra\footnote{We thank D.V. Millionshchikov who drew our attention to this circumstance.} (see \cite {Rinehart}).
We call this algebra the characteristic Lie algebra of the system of equations (\ref{eq2_1}) along the direction $ y $. It is well known (see \cite {ZMHS-UMJ, ZMHSbook}) that the function $ I $ is a $ y $ -integral of the system (\ref{eq2_1}) if and only if it belongs to the kernel of each operator from $ \mathcal {L } (y, N) $. Since the $ y $ -integral depends only on a finite number of dynamic variables, we can use the well-known Jacobi theorem on the existence of a nontrivial solution of a system of first-order linear differential equations with one unknown function. From this theorem it is easy to deduce that in the Darboux integrable case in the algebra $ \mathcal {L} (y, N) $ there must exist a finite basis $ Z_1, Z_2, ... Z_k $, consisting of linearly independent operators such that any element $ Z $ of $ \mathcal {L} (y, N) $ can be represented as a linear combination $ Z = a_1Z_1 + a_2Z_2 + ... a_kZ_k $, where the coefficients $ a_1, a_2 ,. ..a_k $ are analytic functions of dynamical variables defined in some open set. Moreover, from the equality $ a_1Z_1 + a_2Z_2 + ... a_kZ_k = 0 $ it follows that $ a_1 = a_2 = ... = a_k = 0 $. In this case, we call the algebra $ \mathcal {L} (y, N) $ finite-dimensional. Similarly, we can define the characteristic algebra $ \mathcal {L} (x, N) $ in the direction $ x $. It is clear that the system (\ref{eq2_1}) is Darboux integrable if and only if the characteristic algebras in both directions are finite-dimensional.

For the sake of convenience, we introduce the notation $ {\rm ad} _X (Z): = \left [X, Z \right] $. We note that in our study the operator $ {\rm ad} _ {D_x} $ plays a key role. Below we shall apply $ D_x $ to smooth functions depending on dynamical variables
  $ {\bf u}, {\bf u} _x, {\bf u} _ {xx}, \ldots $. As was shown above, the operators $ D_y $ and $ Y $ coincide on this class of functions.
Therefore, the equality $ \left [D_x, D_y \right] = 0 $ immediately implies $ \left [D_x, Y \right] = 0 $. Replacing $ Y $ by virtue of (\ref{eq2_5}), and collecting in the resulting relation the coefficients of the independent variables $ \left\{u_ {i, y}
\right\}_{i=0}^N$, we obtain
\begin{equation}
\left[ D_x, Y_i \right] = -a_i Y_i, \quad {\rm{忍忱快}} \quad a_i = \alpha_i u_{i,x} + \gamma_i.  \label{eq2_8} 
\end{equation}
It is clear that the operator $ {\rm ad} _ {D_x} $ takes the characteristic Lie algebra into itself. We describe the kernel of this mapping:

\begin{lemma} \cite{ZhMuk91, Sh1995, ZMHSbook} \label{lemma1}
If the vector field of the form
\begin{equation} \label{eq2_9}
Z = \sum_i z_{1,i} \frac{\partial}{\partial u_{i,x}} + z_{2,i} \frac{\partial}{\partial	u_{i,xx}} + \cdots
\end{equation}
solves the equation $\left[ D_x, Z \right] = 0$, then $Z=0$.
\end{lemma}

\section{Method of test sequences}

We call the sequence of operators $ W_0, W_1, W_2, \ldots $ in the algebra $ \mathcal {L} (y, N) $ a test sequence if $ \forall $ $ m $ holds:
\begin{equation}  \label{eq3_1} 
\left[D_x, W_m\right] = \sum_{j=0} ^m w_{j,m}W_j.
\end{equation}
The test sequence allows us to derive the integrability conditions for a system of hyperbolic type
(\ref{eq2_1}) (see \cite{H2007, ZMHSbook, ZMHS-UMJ}). Indeed, assume that (\ref{eq2_1}) is Darboux integrable. Then among the operators $ W_0, W_1, W_2, \ldots $ there is only a finite set of linearly independent elements through which all the others are expressed.  In other words, there exists an integer $ k $ such that the operators $ W_0, \ldots W_k $ are linearly independent and $ W_ {k + 1} $ is expressed as follows:
\begin{equation}   \label{eq3_2} 
W_{k+1} =\lambda_k W_k + \cdots + \lambda_0 W_0.
\end{equation}
We apply the operator $ {\rm ad} _ {D_x} $ to both sides of the equality (\ref{eq3_2}). As a result, we obtain the relation
\begin{eqnarray} 
\fl   \sum_{j=0} ^k w_{j,k+1}  W_j + w_{k+1,k+1}  \sum_{j=0} ^k \lambda_j W_j = \sum_{j=0} ^k D_x(\lambda_j)W_j + \nonumber\\
+ \lambda_k \sum_{j=0} ^k w_{j,k} W_j  + \lambda_{k-1}  \sum_{j=0} ^{k-1}  w_{j,k-1} W_j+ \cdots+ \lambda_0 w_{0,0} W_0. \label{eq3_3} 
\end{eqnarray}
Collecting coefficients for independent operators, we obtain a system of differential equations for the coefficients $ \lambda_0, \lambda_1, \ldots \lambda_k $. The resulting system is overdetermined, since $ \lambda_j $ is a function of a finite number of dynamical variables $ {\bf u}, {\bf u} _x, \ldots $. The consistency conditions for this system define the integrability conditions for  (\ref{eq2_1}). For example, collecting the coefficients for $ W_k $ we get the first equation of the indicated system:
\begin{equation} \label{eq3_4} 
D_x(\lambda_k) = \lambda_k (w_{k+1,k+1}  - w_{k,k})+ w_{k,k+1},
\end{equation}
which is also overdetermined.

Below in this section, we use two test sequences to refine the form of the functions $ \alpha_n $, $ \beta_n $, $ \gamma_n $.

\subsection{First test sequence}

Let us define a sequence of operators in the characteristic algebra $ \mathcal {L} (y, N) $ by the following recurrence formula:
\begin{equation}
Y_0, \quad Y_1, \quad W_1 = \left[ Y_0, Y_1 \right], \quad W_2 = \left[Y_0, W_1 \right], \ldots W_{k+1} =\left[ Y_0, W_k \right], \ldots  \label{eq3_5} 
\end{equation}
Above (see (\ref{eq2_8}), the commutation relations for the first two terms of this sequence were derived:
\begin{eqnarray}  \label{eq3_6} 
\fl \left[ D_x, Y_0 \right] = -a_0 Y_0 = -(\alpha_0 u_{0,x}+\gamma_0)Y_0, \quad \left[D_x, Y_1 \right] = -a_1 Y_1=-(\alpha_1 u_{1,x} + \gamma_1 )Y_1.
\end{eqnarray}
Applying the Jacobi identity and using the last formulas, we derive:
\begin{equation}  \label{eq3_7} 
\left[D_x, W_1 \right] = -(a_0+a_1)W_1 - Y_0(a_1)Y_1 + Y_1 (a_0)Y_0.
\end{equation}
We can prove by induction that (\ref{eq3_5}) is a test sequence. Moreover, for any $ k \geq 2 $ the following formula
\begin{equation} \label{eq3_8} 
\left[ D_x, W_k\right] = p_k W_k + q_k W_{k-1}  + \cdots,
\end{equation}
holds where the functions $p_k$, $q_k$ are found due to the rule
\begin{equation}
p_k = -(a_1 + k a_0), \quad q_k = \frac{k-k^2} {2}  Y_0(a_0)- Y_0(a_1)k.  \label{eq3_9}
\end{equation}
By assumption, in the algebra $ \mathcal {L} (y, N) $ there exists only a finite set of linearly independent elements of the sequence (\ref{eq3_5}). Hence, there exists a natural  $ M $ such that:
\begin{equation} \label{eq3_10} 
W_M = \lambda W_{M-1} + \cdots,
\end{equation}
operators $ Y_0, Y_1, W_1, \ldots, W_ {M-1} $ are linearly independent, and three dots stand for a linear combination of the operators $ Y_0, Y_1, W_1, \ldots, W_ {M-2} $.

\begin{lemma} \label{lemma2} 
The operators $ Y_0, Y_1, W_1 $ are linearly independent.
\end{lemma}

Proof. Let us assume the contrary. Suppose that
\begin{equation} \label{eq3_11} 
\lambda_1 W_1 + \mu_1 Y_1 + \mu_0 Y_0 = 0.
\end{equation}
The operators $ Y_0 $, $ Y_1 $ have the form $ Y_0 = \frac {\partial} {\partial u_0} + \cdots $, $ Y_1 = \frac {\partial} {\partial u_1} + \cdots $ while $ W_1 $ does not contain terms of the form $ \frac {\partial} {\partial u_0} $ and $ \frac {\partial} {\partial u_1} $, hence the coefficients $ \mu_1, \mu_0 $ are zero. If, in addition, $ \lambda_1 \neq 0 $, then $ W_1 = 0 $. We apply the operator $ {\rm ad} _ {D_x} $ to both sides of the last equality, then by
(\ref{eq3_7}) we obtain the equation
\begin{equation*}
Y_0(a_1)Y_1 - Y_1(a_0)Y_0 = 0
\end{equation*}
which implies:  $Y_0(a_1)= \alpha_{1,u_0} u_{1,x} + \gamma_{1,u_0} = 0$ and $Y_1(a_0)=\alpha_{0,u_1} u_{0,x} + \gamma_{0,u_1} =0$. By virtue of the independence of the variables $ u_ {0, x} $ and $ u_ {1, x} $, we obtain that $ \alpha_ {1, u_0} = \alpha_ {0, u_1} = 0 $. But this contradicts the assumption of (\ref{eq11}) that $ \frac {\partial \alpha (u_ {n + 1}, u_n, u_ {n-1})} {\partial u_ {n \pm 1}} \neq 0 $. The proof is complete. 

\begin{lemma}\label{lemma33}
If the expansion of the form (\ref{eq3_10}) holds, then
\begin{equation} \label{eq3_20} 
\alpha(u_1,u_0,u_{-1}) = \frac{P'(u_0)}{P(u_0)+Q(u_{-1} )} + \frac{1} {M-1}  \frac{Q'(u_0)}{P(u_1)+Q(u_0)} - c_1(u_0).
\end{equation}
\end{lemma}

Proof.
It is not difficult to show that equation (\ref{eq3_4}) for the sequence (\ref{eq3_5}) has the form:
\begin{equation}
D_x (\lambda) = -a_0 \lambda - \frac{M(M-1)}{2}  Y_0(a_0)-M Y_0(a_1).  \label{eq3_12} 
\end{equation}
We simplify the relation (\ref{eq3_12}) using formulas
\begin{eqnarray}
&&Y_0(a_0)= \left(\frac{\partial}{\partial u_0}  + (\alpha_0 u_{0,x}+\gamma_0) \frac{\partial}{\partial u_{0,x}} \right)(\alpha_0 u_{0,x}+\gamma_0) = \nonumber\\
&&\hphantom{Y_0(a_0)}=\left(\alpha_{0, u_0}  + \alpha_0^2\right)u_{0x} +\gamma_{0,u_0}+\alpha_0 \gamma_0 ,\\ \label{eq3_13} 
&&Y_0(a_1)= \alpha_{1,u_0}  u_{1,x} + \gamma_{1,u_0}. \nonumber
\end{eqnarray}
A simple analysis of the equation (\ref{eq3_12}) shows that $ \lambda = \lambda (u_0, u_1) $. Therefore, (\ref{eq3_12}) is rewritten as	
\begin{eqnarray*}
\fl \lambda_{u_0}  u_{0,x} + \lambda_{u_1}  u_{1,x} = -\left( (\alpha_0 \lambda + \frac{M(M-1)}{2} (\alpha_{0,u_0} +\alpha^2_0)\right)u_{0,x} - M \alpha_{1,u_0}  u_{1,x} - \\
-\left(\gamma_0 \lambda + \frac{M(M-1)}{2}(\gamma_{0,u_0}+\alpha_0 \gamma_0 ) + M \gamma_{1,u_0} \right).
\end{eqnarray*}
Collecting the coefficients in front of the independent variables $ u_ {0, x} $, $ u_ {1, x} $, we derive an overdetermined system of differential equations in $ \lambda $:
\begin{eqnarray} 
&&\lambda_{u_0}  = -\alpha_0 \lambda - \frac{M(M-1)}{2} (\alpha_{0, u_0}  + \alpha^2_0), \quad \lambda_{u_1}  = -M \alpha_{1,u_0},\label{eq3_14} \\
&&\gamma_0 \lambda + \frac{M(M-1)}{2}(\gamma_{0,u_0}+\alpha_0 \gamma_0 ) + M \gamma_{1,u_0} = 0. \label{eq3_14p} 
\end{eqnarray}
Note that equations (\ref{eq3_14}) do not contain function $ \gamma $ and completely coincide with the equations studied in our article \cite{HabPoptsovaSIGMA17}. Lemma 3 immediately follows from Lemma 3.2 in \cite{HabPoptsovaSIGMA17}. In what follows we use the equation (\ref{eq3_14p}) to refine the function $ \gamma $.

\subsection{Second test sequence}

We construct a test sequence containing the operators $ Y_0 $, $ Y_1 $, $ Y_2 $ and their multiple commutators:
\begin{eqnarray}
Z_0 = Y_0, \, Z_1 = Y_1, \, Z_2 = Y_2, \, Z_3 = \left[Y_1, Y_0\right], \, Z_4 = \left[Y_2, Y_1\right], \nonumber\\ Z_5 = \left[Y_2, Z_3 \right], \, Z_6 = \left[ Y_1, Z_3\right], \, Z_7 = \left[Y_1, Z_4 \right], \, Z_8 = \left[Y_1, Z_5 \right].  \label{eq3_21} 
\end{eqnarray}
Elements of the sequence $ Z_m $ for $ m> 8 $ are determined by the recurrence formula $ Z_ {m} = \left [Y_1, Z_ {m-3} \right] $. Note that this is the simplest test sequence generated by iterations of the map $ Z \rightarrow \left [Y_1, Z \right] $, which contains the operator $ \left [Y_2, \left [Y_1, Y_0 \right] \right] = Z_5 $.

\begin{lemma} \label{lemma3}
The operators $ Z_0, \, Z_1, \ldots Z_5 $ are linearly independent.
\end{lemma}

Proof.
Arguing as in the proof of Lemma 1, we can verify that the operators $ Z_0, \, Z_1, \ldots Z_4 $ are linearly independent. Let us prove the lemma~\ref {lemma3} by contradiction. Let's assume that
\begin{equation}  \label{eq3_22} 
Z_5 = \sum_{j=0} ^4 \lambda_j Z_j.
\end{equation}
First we derive the formulas by which the operator $ {\rm ad} _ {D_x} $ acts on the operators $ Z_i $. For $ i = 0,1,2 $, they are immediately obtained from relation
\begin{equation*}
\left[D_x, Y_i \right] = -a_i Y_i.
\end{equation*}
Recall that $a_i =\alpha_i u_{i,x}+\gamma_i= \alpha(u_{i-1},u_i, u_{i+1})u_{i,x}+\gamma(u_{i-1},u_i, u_{i+1})$. For $i = 3,4,5$ we have
\begin{eqnarray*}
&&\left[ D_x, Z_3 \right] = -(a_1 + a_0) Z_3 + \cdots,\\
&&\left[D_x, Z_4 \right] = -(a_2 + a_1)Z_4 + \cdots,\\
&&\left[D_x, Z_5 \right] = -(a_0 + a_1 + a_2) Z_5 + Y_0(a_1) Z_4 - Y_2(a_1) Z_3 + \cdots	
\end{eqnarray*} 
By applying the operator ${\rm ad}_{D_x}$ to both sides of (\ref{eq3_22}), we obtain
\begin{eqnarray}
\fl  -(a_0 + a_1 + a_2) (\lambda_4 Z_4 + \lambda_3 Z_3 + \cdots) + Y_0(a_1) Z_4  - Y_2(a_1) Z_3 + \cdots =\nonumber\\
= \lambda_{4,x} Z_4 + \lambda_{3,x} Z_3 - \lambda_4 (a_1 + a_2) Z_4 -\lambda_3 (a_0 + a_1) Z_3  + \cdots \label{eq3_23} 
\end{eqnarray}
Collecting the coefficients for $ Z_4 $ in the equality (\ref{eq3_23}), we obtain the following equation:
\begin{equation}  \label{eq3_24} 
\lambda_{4,x} = -(\alpha_0 u_{0,x} + \gamma_0) \lambda_4 -(\alpha_{1,u_0}  u_{1,x}+\gamma_{1,u_0}).
\end{equation}
A simple analysis of the equation (\ref{eq3_24}) shows that $ \lambda = \lambda (u_0, u_1) $. Consequently, $\lambda_{4,x}= \lambda_ {4, u_0}u_{0,x}+\lambda_ {4, u_1}u_{1,x}$ and  equation (\ref{eq3_24}) reduces to a system of three equations $ \gamma_0 \lambda_4 + \gamma_ {1, u_0} = 0 $, $ \lambda_ {4, u_0} = - \alpha_0 \lambda_4 $ and $ \lambda_ { 4, u_1} = - \alpha_ {1, u_0} $. From these equations it follows that $ \lambda_4 = 0 $. Otherwise, if $ \lambda_4 \neq 0 $, then $ \alpha_0 = - \left (\log \lambda_4 \right) _ {u_0} $, which implies that $ (\alpha_0) _ {u_ {- 1}} = 0 $ and this contradicts the requirement that $ \alpha (u_1, u_0, u_ {- 1}) $ essentially depends on $ u_1 $ and $ u_ {-l} $, hence $ \lambda_4 = 0 $. Then from (\ref{eq3_24}) we have $ \alpha_ {1, u_0} = 0 $, which again leads to a contradiction.

Let's return to the sequence (\ref{eq3_21}). For further work, it is necessary to describe the action of the operator $ {\rm ad} _ {D_x} $ on all elements of this sequence. It is convenient to separate the sequence (\ref{eq3_21}) into three subsequences $ \{Z_ {3m} \} $, $ \{Z_ {3m + 1} \} $ and $ \{Z_ {3m + 2} \} $.

\begin{lemma} \label{lemma4}
The action of the operator $ {\rm ad} _ {D_x} $ on the sequence (\ref{eq3_21}) is given by the following formulas:
\begin{eqnarray*}
\fl \left[D_x, Z_{3m} \right] = - (a_0 + m a_1)Z_{3m} +\left( \frac{m-m^2} {2}  Y_1 (a_1)- m Y_1 (a_0)\right)Z_{3m-3} + \cdots,\\
\fl \left[ D_x, Z_{3m+1} \right] = -(a_2 + m a_1)Z_{3m+1}  + \left( \frac{m-m^2} {2}  Y_1 (a_1)- m Y_1 (a_2)\right) Z_{3m-2}  + \cdots,\\
\fl \left[D_x, Z_{3m+2}  \right] = - (a_0 + m a_1 + a_2)Z_{3m+2}  +Y_0 (a_1)Z_{3m+1} + Y_2 (a_1)Z_{3m} -\\
\hphantom{\left[D_x, Z_{3m+2}  \right] =}  - (m-1) \left( \frac{m}{2}  Y_1 (a_1)+ Y_1 (a_0 + a_2) \right)Z_{3m-1}  + \cdots
\end{eqnarray*}
\end{lemma}
Lemma~\ref{lemma4} is easily proved by induction.

\begin{theorem} \label{theorem2}
Assume that the operator $ Z_ {3k + 2} $ is represented as a linear combination
\begin{equation} \label{eq3_25} 
Z_{3k+2}  = \lambda_k Z_{3k+1}  + \mu_k Z_{3k} + \nu_k Z_{3k-1}  + \cdots
\end{equation}
of the previous members of the sequence (\ref{eq3_21}) and none of the operators $ Z_ {3j + 2} $ for $ j <k $ is a linear combination of the operators $ Z_s $ with $ s <3j + 2 $. Then the coefficient $ \nu_k $ satisfies the equation
\begin{equation} \label{eq3_26} 
D_x(\nu_k) = -a_1 \nu_k - \frac{k(k-1)}{2}  Y_1 (a_1)- (k-1)Y_1 (a_0 + a_2).
\end{equation}
\end{theorem}

\begin{lemma} \label{lemma5} 
Assume that all the conditions of Theorem~\ref {theorem2} are satisfied. Suppose that the operator $ Z_ {3k} $ (the operator $ Z_ {3k + 1} $) is linearly expressed in terms of the operators $ Z_i $, $ i <3k $. Then in this expansion the coefficient at $ Z_ {3k-1} $ is zero.
\end{lemma}

Proof. Let us prove the assertion  by contradiction, assume that in formula
\begin{equation}  \label{eq3_27} 
Z_{3k} = \lambda Z_{3k-1}  + \cdots
\end{equation}
the coefficient $ \lambda $ is nonzero. We apply the operator $ {\rm ad} _ {D_x} $ to both sides of the equation (\ref{eq3_27}). As a result, according to Lemma~\ref{lemma4}, we get:
\begin{equation} 
  -(a_0 + k a_1)\lambda Z_{3k-1}  + \cdots 
= D_x(\lambda) Z_{3k-1}  - \lambda (a_0 + (k-1)a_1 + a_2)Z_{3k-1}  + \cdots \label{eq3_28} 
\end{equation}
Collecting the coefficients at $ Z_ {3k-1} $, we obtain that the coefficient $ \lambda $ should satisfy the equation
\begin{equation*}
D_x(\lambda) = \lambda (a_2 - a_1).
\end{equation*}
According to our assumption above, $ \lambda $ does not vanish and, therefore,
\begin{equation} \label{eq3_29}
D_x(\log \lambda) = a_2 - a_1.
\end{equation}
Since $ \lambda $ depends on a finite set of dynamical variables, according to the equation (\ref{eq3_29}) $ \lambda $ can depend only on $ u_1 $ and $ u_2 $. Therefore, from (\ref{eq3_28}) we get that
\begin{equation*}
(\log \lambda)_{u_1}  u_{1,x} + (\log \lambda)_{u_2}  u_{2,x} = \alpha_2 u_{2,x} +\gamma_2 - \alpha_1 u_{1,x}-\gamma_1.
\end{equation*}
The variables $ u_ {1, x} $, $ u_ {2, x} $ are independent, so the last equation is equivalent to the system of equations $ \alpha_1 = - (\log \lambda) _ {u_1} $, $ \ alpha_2 = ( \log \lambda) _ {u_2} $, $ \gamma_2- \gamma_1 = 0 $. Consequently, $ \alpha_1 = \alpha_1 (u_1, u_2) $ depends only on $ u_1 $, $ u_2 $. The latter contradicts the assumption that $ \alpha_1 $ essentially depends on $ u_0 $. The contradiction shows that the assumption $ \lambda \neq 0 $ is false. The lemma is proved.

In order to prove Theorem~\ref{theorem2}, we apply the operator $ {\rm ad} _ {D_x} $ to both sides of the equality (\ref{eq3_25}) and simplify due to the formulas in Lemma~\ref{lemma4}. Collecting the coefficients for $ Z_ {3k-1} $, we obtain the equation (\ref{eq3_26}).

Let us find the exact values of the coefficients of equation (\ref{eq3_26})
\begin{eqnarray*}
&&Y_1(a_0) = Y_1 (\alpha_0 u_{0,x} + \gamma_0)= \alpha_{0, u_1}  u_{0,x} + \gamma_{0,u_1}, \\
&&Y_1(a_2) =  Y_1 (\alpha_2 u_{2,x} + \gamma_2)= \alpha_{2,u_1}  u_{2,x} + \gamma_{2,u_1}, \\
&&Y_1(a_1) = Y_1(\alpha_1 u_{1,x}+\gamma_1)= (\alpha_{1, u_1} + \alpha^2_1) u_{1,x} + \gamma_{1,u_1} + \gamma_1 \alpha_1.
\end{eqnarray*}
and substitute them into (\ref{eq3_26}):
\begin{eqnarray}
\fl  D_x(\nu_k) = -(\alpha_1 u_{1,x}+\gamma_1)\nu_k  - \frac{k(k-1)}{2} \left( (\alpha_{1,u_{1} }+ \alpha_1^2) u_{1,x} + \gamma_{1,u_1} + \gamma_1 \alpha_1\right)-\nonumber\\
  - (k-1)(\alpha_{0,u_1}  u_{0,x} + \alpha_{2,u_1}  u_{2,x} + \gamma_{0,u_1} + \gamma_{2,u_1} ). \label{eq3_30} 
\end{eqnarray}
A simple analysis of the equation (\ref{eq3_30}) shows that $ \nu_k $ can  depend only on the variables $ u_0, \, u_1, \, u_2 $. Consequently,
\begin{equation}  \label{eq3_31} 
D_x(\nu_k) = \nu_{k,u_0}  u_{0,x} + \nu_{k,u_1}  u_{1,x} + \nu_{k, u_2}  u_{2,x}.
\end{equation}
Substituting (\ref{eq3_31}) in (\ref{eq3_30}) and collecting coefficients for independent variables, we obtain a system of equations by the coefficient $ \nu_k $:
\begin{eqnarray}
&&\nu_{k,u_0}  = -(k-1)\alpha_{0,u_1} , \label{eq3_32} \\
&&\nu_{k,u_1}  = -\alpha_1 \nu_k  - \frac{k(k-1)}{2}  (\alpha_{1,u_1}  + \alpha^2_1), \label{eq3_33} \\
&&\nu_{k,u_2}  = -(k-1)\alpha_{2,u_1}, \label{eq3_34} \\
&&0 = \gamma_1 \nu_k  + \frac{k(k-1)}{2} \left( \gamma_{1,u_1} + \gamma_1 \alpha_1\right) + (k-1)(\gamma_{0,u_1} + \gamma_{2,u_1}).
\end{eqnarray}
Substituting the expression for the function $ \alpha $, given by the formula (\ref{eq3_20}) into the equation (\ref{eq3_32}), we get
\begin{equation*}
\nu_{k,u_0} = \frac{k-1} {M-1}  \frac{P'(u_1) Q'(u_0)}{(P(u_1)+ Q(u_0))^2} .
\end{equation*}
We integrate the last equation with respect to the variable $ u_0 $
\begin{equation} \label{eq3_35} 
\nu_k = -\frac{k-1} {M-1}  \frac{P'(u_1)}{P(u_1) + Q(u_0)} + H(u_1,u_2).
\end{equation}
Since $ \nu_ {k, u_2} = H_ {u_2} $, the equation (\ref{eq3_34}) is rewritten as
\begin{equation*}
H_{u_2} =(k-1)\frac{P'(u_2)Q'(u_1)} {(P(u_2)+Q(u_1))^2}.
\end{equation*}
Integrating the latter, we obtain an exact expression for the function $ H $
\begin{equation*}
H = -(k-1) \left( \frac{Q'(u_1)}{P(u_2)+Q(u_1)} + A(u_1) \right),
\end{equation*}
which gives
\begin{equation} \label{eq3_36} 
\nu_k = -(k-1) \left(\frac{1} {M-1}  \frac{P'(u_1)}{P(u_1) + Q(u_0)} + \frac{Q'(u_1)}{P(u_2) + Q(u_1)} + A(u_1) \right).
\end{equation}
We substitute the found expressions for the functions $ \alpha $ and $ \nu_k $ into the equation (\ref{eq3_33})
\begin{eqnarray}
  -\frac{(k-1)}{M-1}  \left( \frac{P''(u_1)}{P(u_1)+Q(u_0)} - \frac{P'^2(u_1)}{(P(u_1)+Q(u_0))^2} \right)-\nonumber\\
-(k-1)\left(\frac{Q''(u_1)}{P(u_2)+Q(u_1)} - \frac{Q'^2(u_1)}{(P(u_2)+Q(u_1))^2} + A'(u_1) \right)= \nonumber\\
=(k-1)\left(\frac{P'(u_1)}{P(u_1)+Q(u_0)} + \frac{1} {M-1}  \frac{Q'(u_1)}{P(u_2)+Q(u_1)} -c_1(u_1) \right) \times\nonumber\\
\times \left( \frac{1} {M-1}  \frac{P'(u_1)}{P(u_1)+Q(u_0)}+ \frac{Q'(u_1)}{P(u_2)+Q(u_1)} + A(u_1)\right)- \nonumber\\
- \frac{k(k-1)}{2}  \left( \frac{P''(u_1)}{P(u_1)+Q(u_0)} + \frac{1} {M-1} \frac{Q''(u_1)}{P(u_2)+Q(u_1)} -\right.\nonumber\\
- \frac{1} {M-1}  \frac{Q'^2(u_1)}{(P(u_2)+Q(u_1))^2}  + \frac{1} {M-1}  \frac{2 Q'(u_1) P'(u_1)}{(P(u_1)+Q(u_0))(P(u_2)+Q(u_1))} +\nonumber\\
\left. + \frac{1} {(M-1)^2} \frac{Q'^2(u_1)}{(P(u_2)+Q(u_1))^2} -c'_1(u_1)-\right.\nonumber\\
\left. - 2 c_1(u_1)\left( \frac{P'(u_1)}{P(u_1)+Q(u_0)} + \frac{1} {M-1}  \frac{Q'(u_1)}{P(u_2)+Q(u_1)} \right) + c^2_1(u_1) \right). \label{eq3_37} 
\end{eqnarray}
Obviously, according to the assumption $ \frac {\partial} {\partial u_1} \alpha (u_1, u_0, u_ {-1}) \neq 0 $, $ \frac {\partial} {\partial u_ {-l} } \alpha (u_1, u_0, u_ {- 1}) \neq 0 $ the functions $ P '(u_2) $ and $ Q' (u_0) $ do not vanish. Consequently, the variables
\begin{equation*}
\frac{Q'^2(u_1)}{(P(u_2)+Q(u_1))^2} , \quad \frac{P'^2(u_1)}{(P(u_1)+Q(u_0))^2} , \quad \frac{P'(u_1)Q'(u_1)}{(P(u_1)+Q(u_0))(P(u_2)+Q(u_1))}
\end{equation*}
are independent. Collecting the coefficients of these variables in (\ref{eq3_37}), we obtain a system of two equations
\begin{equation} \label{eq3_38} 
\left( 1 - \frac{1} {M-1} \right) \left(1 - \frac{k}{2(M-1)} \right) = 0, \quad 1+ \frac{1} {(M-1)^2}  = \frac{k}{M-1} .
\end{equation}
The system (\ref{eq3_38}) has two solutions: $ M = 0, \, k = -2 $ and $ M = 2, \, k = 2 $. Since $ k $ must be greater than zero, we have $ M = 2, \, k = 2 $. The last argument completes the proof of the theorem~\ref{theorem2}.

Thus, we have proved that $ M = 2 $, $ k = 2 $. Expansions
(\ref{eq3_10}) 我 (\ref{eq3_25}) take the form
\begin{equation} \label{eq4_2}
W_2 = \lambda W_1 + \sigma Y_1 + \delta Y_0,
\end{equation}
\begin{equation} \label{eq4_13}
Z_8 = \lambda Z_7 + \mu Z_6 + \nu Z_5 + \rho Z_4 + \kappa Z_3 + \sigma Z_2 + \delta Z_1 + \eta Z_0.
\end{equation}
The following is valid
\begin{theorem}
The expansions (\ref{eq4_2}), (\ref{eq4_13}) take place if and only if the functions $ \alpha $, $ \gamma $ in the equation (\ref{eq1}) have the form:
\begin{eqnarray}
&&\alpha(u_{n+1},u_n,u_{n-1}) = \frac{1}{u_n - u_{n-1}} - \frac{1}{u_{n+1}-u_n},\\
&&\gamma(u_{n+1},u_n,u_{n-1}) = r'(u_n) - r(u_n)\alpha(u_{n+1},u_n,u_{n-1}), \label{gamma_p3}
\end{eqnarray}
where $r(u_n) = \frac{k_1}{2} u^2_n + k_2 u_n + k_3$ and the factors $k_i$ -- are arbitrary constants.
\end{theorem}

Proof.
Consider the relation (\ref{eq4_2}). Using the relations (\ref{eq3_6}), (\ref{eq3_7}) and applying the Jacobi identity, we get
\begin{eqnarray} 
\fl  \left[ D_x, W_2 \right] = -(2 a_0 + a_1)W_2 - Y_0(a_0 + 2a_1)W_1 +\nonumber\\
 +(2 Y_0 Y_1 (a_0) - Y_1 Y_0 (a_0)) Y_0 - Y_0 Y_0 (a_1)Y_1. \label{eq4_1}
\end{eqnarray}
It is obvious that only one term in the formula (\ref{eq4_2}) contains the operator of differentiation $ \frac {\partial} {\partial u_1} $, namely $ \sigma Y_1 $, and only one term contains $ \frac {\partial} {\partial u_0} $, namely $ \sigma Y_0 $. Consequently, $ \sigma = 0 $, $ \delta = 0 $ and the expansion of (\ref{eq4_2}) takes the form
\begin{equation*} 
W_2 = \lambda W_1.
\end{equation*}
Applying the operator $ {\rm ad} _ {D_x} $ to both sides of the last relation, we obtain
\begin{eqnarray*}
\fl  -(2 a_0 + a_1)W_2 - Y_0(a_0 + 2a_1)W_1 + (2 Y_0 Y_1 (a_0) - Y_1 Y_0 (a_0)) Y_0 - Y_0 Y_0 (a_1)Y_1 =\\
=D_x(\lambda) W_1 + \lambda \left( -(a_0 + a_1) W_1 + Y_1(a_0) Y_0  - Y_0(a_1)Y_1\right).
\end{eqnarray*}
Collecting the coefficients for the operators $ W_2 $, $ W_1 $, $ Y_1 $, $ Y_0 $, we arrive at the following system:
\begin{eqnarray}
&&D_x(\lambda) = -a_0 \lambda - Y_0 (a_0 + 2 a_1), \label{eq4_3} \\
&&-Y_0 Y_0 (a_1) = -\lambda Y_0 (a_1) , \label{eq4_4} \\
&&2 Y_0 Y_1 (a_0) - Y_1 Y_0 (a_0) = \lambda Y_1(a_0). \label{eq4_5} 
\end{eqnarray} 
Examining the first equation of the obtained system, we observe that $ \lambda = \lambda (u_0, u_1) $ and then simplifying all the equations, we arrive at the following system:
\begin{eqnarray}
&&\lambda_{u_0} = -\alpha_0 \lambda - (\alpha_{0,u_0}  + \alpha^2_0),\label{eq4_5_0} \\
&&\lambda_{u_1}  = - 2 \alpha_{1,u_0} , \label{eq4_5_1} \\
&&\alpha_{1,u_0 u_0} =\lambda \alpha_{1,u_0}, \quad \alpha_{0,u_0u_1}   = \lambda \alpha_{0,u_1} . \label{eq4_8}
\end{eqnarray}
\begin{eqnarray}
&&\gamma_0 \lambda + \gamma_{0,u_0} + \gamma_0 \alpha_0 + 2 \gamma_{1,u_0} = 0, \label{eq4_g1}\\
&&\gamma_{1,u_0u_0} = \lambda \gamma_{1,u_0}, \label{eq4_g2}\\
&&\gamma_{0,u_0u_1} + \gamma_0 \alpha_{0,u_1} - \gamma_{0,u_1}\alpha_0 = \lambda \gamma_{0,u_1}. \label{eq4_g3}
\end{eqnarray} 
We note at once that the equations (\ref{eq4_5_0})-(\ref{eq4_8}) will be used to refine the functions $ \alpha $ and $ \lambda $, and the equations (\ref{eq4_g1})-(\ref{eq4_g3}) to specify the function $ \gamma $, substituting the already found expression for $\alpha$.

Next, we turn to the decomposition (\ref{eq4_13}).
Putting $ k = 2 $ in the formulas of Lemma~\ref {lemma4}, we obtain
\begin{eqnarray}
\fl \left[D_x, Z_6 \right] = -(\alpha_0 u_{0,x}+2 \alpha_1 u_{1,x})Z_6 + \cdots,\label{eq4_16} \\
\fl  \left[ D_x, Z_7 \right] = -(\alpha_2 u_{2,x}+2 \alpha_1 u_{1,x})Z_7 - (Y_1(\alpha_1 u_{1,x})+2Y_1(\alpha_2 u_{2,x}))Z_4 +\cdots,\label{eq4_17} \\
\fl \left[ D_x, Z_8 \right] = - (\alpha_0 u_{0,x} + 2 \alpha_1 u_{1,x} + \alpha_2 u_{2,x})Z_8 + Y_0(\alpha_1 u_{1,x})Z_7 + Y_2(\alpha_1 u_{1,x})Z_6 -\nonumber\\
 \hphantom{\left[ D_x, Z_8 \right] =}  - \left(   Y_1 (\alpha_1 u_{1,x})+ Y_1 (\alpha_0 u_{0,x} + \alpha_2 u_{2,x})\right) Z_5 + \cdots.\label{eq4_18} 
\end{eqnarray}
Then we apply the operator $ {\rm ad} _ {D_x} $ to both parts of the relation (\ref{eq4_13}) and simplify the resulting equation using (\ref{eq4_16}), (\ref{eq4_17}), (\ref{eq4_18}). Comparison of coefficients at
$ Z_7 $ and $ Z_6 $ gives $ \lambda = 0 $ and $ \mu = 0 $. Thus, the formula (\ref{eq4_13}) is simplified:
\begin{equation} \label{eq4_19}
Z_8 = \nu Z_5 + \rho Z_4 + \kappa Z_3 + \sigma Z_2 + \delta Z_1 + \eta Z_0.
\end{equation}
The following commutation relations hold:
\begin{eqnarray}
\fl \left[ D_x, Z_8 \right] = -(a_2 + 2a_1 + a_0)Z_8 + Y_0(a_1) Z_7-  Y_2(a_1)Z_6 
- Y_1(a_2 + a_1 + a_0) Z_5 +\nonumber \\
\fl \hphantom{\left[ D_x, Z_8 \right] =}  + Y_1 Y_0 (a_1) Z_4  - Y_1 Y_2 (a_1) Z_3 + (Y_1 Y_2 Y_0 (a_1) +Z_5(a_1))Z_1 ,\label{eq4_20}  \\
\fl \left[ D_x, Z_5 \right] = -(a_0 + a_1 + a_2) Z_5 + Y_0(a_1)Z_4 - Y_2(a_1)Z_3 + Y_2 Y_0(a_1) Z_1. \label{eq4_21} 
\end{eqnarray}
We apply $ {\rm ad} _ {D_x} $ to (\ref{eq4_19}), then simplify according to (\ref{eq4_20}), (\ref{eq4_21}), (\ref{eq4_19}) and collect the coefficients for $ Z_5 $
 \begin{equation*}
 -(a_2 + 2 a_1 + a_0) \nu - Y_1(a_2+a_1 + a_0) = D_x(\nu) - (a_2 + a_1 + a_0)\nu
 \end{equation*}
or the same
\begin{equation} \label{I45} 
 D_x(\nu) = -a_1 \nu - Y_1(a_2 + a_1 + a_0).
 \end{equation}
From the equation (\ref{I45}) it follows that $ \nu $ depends on three variables $ \nu = \nu (u, u_1, u_2) $. Thus, the equation (\ref{I45}) reduces to a system of equations: 
\begin{eqnarray}
 && \nu_u = - \alpha_{0,u_1} , \label{I46} \\
 && \nu_{u_1}  = -\alpha_1 \nu - \alpha_{1, u_1}  - \alpha^2_1,  \label{I47} \\
 && \nu_{u_2}  = - \alpha_{2,u_1},  \label{I48} \\
 && \gamma_1 \nu + \gamma_{2,u_1} + \gamma_1 \alpha_1 + \gamma_{1,u_1} + \gamma_{0,u_1} = 0. \label{gamma_2}
 \end{eqnarray}
So, as a result of the investigation of the relations (\ref{eq4_2}), (\ref{eq4_13}), we come to the equations (\ref{eq4_5_0})-(\ref{eq4_8}) and (\ref{I46})-(\ref{I48}), which exactly coincide with the corresponding systems of equations from the work \cite {HabPoptsovaSIGMA17} and, therefore, we get that 
\begin{equation*}
 \alpha(u_{n+1},u_n,u_{n-1}) = \frac{1}{u_n - u_{n-1}} - \frac{1}{u_{n+1}-u_n}.
 \end{equation*}
Using the remaining equations (\ref{eq4_g1})-(\ref{eq4_g3}) and (\ref{gamma_2}) we find $ \gamma $:
\begin{equation*}
\gamma(u_{n+1},u_n,u_{n-1}) = r'(u_n) - r(u_n)\alpha(u_{n+1},u_n,u_{n-1}).
\end{equation*}
It is not difficult to show that the relations (\ref{eq4_2}), (\ref{eq4_13}) take the form:
\begin{eqnarray*}
&&W_2 = \lambda W_1,\quad \lambda = \frac{2}{u_1-u_0},\\
&&Z_8 = \nu Z_5, \quad \nu = -\frac{u_2 - 2 u_1 + u_0}{(u_1-u_0)(u_2-u_1)}.
\end{eqnarray*}

Similarly, we have
\begin{equation}
\beta(u_{n+1},u_n,u_{n-1}) = \tilde{r}'(u_n) - \tilde{r}(u_n)\alpha(u_{n+1},u_n,u_{n-1}),\label{eq3_61}
\end{equation}
where $\tilde{r}(u_n) = \frac{\tilde{k}_1}{2} u^2_n + \tilde{k}_2 u_n + \tilde{k}_3$, and the coefficients $\tilde{k}_i$ -- are arbitrary constants.

The next step of our investigation is to refine the function $ \delta $. To do this, we build a new sequence on a set of multiple commutators.

\section{Specification of the function $\delta$}

Recall that since the right-hand side $ f_i $ of the system (\ref{eq1})  has a special form, the operator $ Y $ can be represented as follows (see (\ref{eq2_5})):
\begin{equation*}
Y=\sum_{i=0}^N u_{i,y} Y_i + R,
\end{equation*}
Here the operator  $R$ is defined by the formula (\ref{op_R}). Consider the following sequence of the operators in the characteristic algebra $\mathcal{L}(y,N)$:
\begin{eqnarray}
Y_{-1}, \, Y_0, \, Y_{1}, \, Y_{0,-1} = \left[Y_{0}, Y_{-1} \right], \, Y_{1,0} = \left[Y_1, Y_0 \right], \label{seq3}\\  
R_0=\left[Y_0, R \right], \, R_1 = \left[Y_0, R_0 \right], \, R_2 = \left[Y_0, R_1 \right], \, \ldots, \, R_{k+1}=\left[ Y_0, R_k \right]. \nonumber
\end{eqnarray}
The following commutation relations hold:
\begin{eqnarray}
&&\left[ D_x, Y_{-1} \right] = -a_{-1} Y_{-1},\quad \left[ D_x, Y_0 \right] = -a_0 Y_0, \quad \left[ D_x, Y_1 \right] = -a_1 Y_1,\label{DxY3}\\
&&\left[ D_x, Y_{1,0} \right] = -(a_0 + a_{1}) Y_{1,0} - Y_{1}(a_0) Y_0 + Y_0(a_{1}) Y_{1}, \label{DxY10} \\
&&\left[ D_x, Y_{0,-1} \right] = -(a_{-1} + a_0) Y_{0,-1} - Y_0(a_{-1}) Y_{-1} + Y_{-1}(a_0) Y_0,  \label{DxY0m1}\\
&&\left[ D_x, R \right] = -\sum_i h_i Y_i,  \label{DxR}
\end{eqnarray}
where $a_i = \alpha_i u_{i,x}+\gamma_i$, $h_i =\beta_i u_{i,x}+\delta_i$.
Using the Jakobi identity and formulas (\ref{DxY3})--(\ref{DxR}) we derive the formulas:
\begin{eqnarray}
 \left[ D_x, R_0 \right] = \left[ D_x, \left[Y_0, R \right] \right]= - \left[Y_0, \left[R, D_x \right]  \right]- \left[ R, \left[ D_x, Y_0\right] \right]= \nonumber\\
=- a_0 R_0 + h_{1} Y_{1,0} - h_{-1} Y_{0,-1} -\nonumber\\
- Y_0(h_{1}) Y_{1} -Y_0(h_{-1})Y_{-1}
+ (R(a_0) - Y_0(h_0))Y_0,\label{S4_eq6}\\
 \left[  D_x, R_1  \right] =  - 2 a_0 R_1 - Y_0(a_0) R_0+ \cdots,\\
 \left[  D_x, R_2  \right] =  - 3 a_0 R_2 - 3Y_0(a_0) R_1- Y^2_0(a_0)R_0+ \cdots, \\
 \left[  D_x, R_3  \right] =  - 4 a_0 R_2 - 6Y_0(a_0) R_1- 4 Y^2_0(a_0)R_1- Y^3_0(a_0)R_0+\cdots,
\end{eqnarray}
where three dots stand for a linear combination of the operators  $Y_{1,0}, Y_{0,-1}, Y_{1}, Y_0, Y_{-1}$. It can be proved by induction that
\begin{equation}
\left[ D_x, R_n \right] = p_n R_n + q_n R_{n-1}+\cdots,
\end{equation}
where
\begin{equation}
p_n = -(n+1)a_0, \quad q_n = - \frac{n^2+n}{2}Y_0(a_0),
\end{equation}
and three dots stand for a linear combination of the operators $R_k$, $k<n-1$ and   $Y_{1,0}, Y_{0,-1}, Y_{1}, Y_0, Y_{-1}$.

Now we consider two different cases:

$i)$ The operator $R_0$ is linearly expressed in terms of the opeartors (\ref{seq3}).

$ii)$ The operator $R_0$ is not linearly expressed in terms of the operators (\ref{seq3}).

Let us focus on the case $i)$. It follows from the formula (\ref{S4_eq6}) that this linear expansion must have the form
\begin{equation} \label{exps_R0}
R_0 = \lambda R + \mu Y_{1,0} + \tilde{\mu} Y_{0,-1} + \nu Y_{1} + \eta Y_0 + \epsilon Y_{-1}.
\end{equation}
The operators in the right hand side of this formula are linearly independent.

By applying the operator ${\rm ad}_{D_x}$ to both sides of (\ref{exps_R0}), we obtain 
\begin{eqnarray}
\fl -a_0 (\lambda R + \mu Y_{1,0} + \tilde{\mu} Y_{0,-1}+\cdots) + h_{1} Y_{1,0} - h_{-1} Y_{0,-1} + \cdots = \nonumber\\
=  D_x(\lambda) R + D_x(\mu) Y_{1,0} + \mu (-(a_0 + a_{1}) Y_{1,0} + \cdots) + \nonumber\\
+D_x(\tilde{\mu}) Y_{0,-1} + \tilde{\mu} (-(a_{-1}+a_0)Y_{0,-1}+\cdots). \label{S4_eq13}
\end{eqnarray}
Three dots stand for a linear combination of the operators $Y_{-1}, Y_0, Y_1$. Collecting coefficients for the independent operators $R$, $Y_{1,0}$, $Y_{0,-1}$, we obtain the system of differential equations for the coefficients $\lambda$, $\mu$, $\tilde{\mu}$
\begin{eqnarray}
&&D_x(\lambda) = -a_0 \lambda, \label{S4_eq13}\\
&&D_x(\mu) = a_{1} \mu +h_{1}, \quad D_x(\tilde{\mu}) = a_{-1} \tilde{\mu} - h_{-1}. \label{S4_eq14}
\end{eqnarray}
The equation (\ref{S4_eq13}) has the form: $D_x(\lambda) = -(\alpha_0 u_{0,x}+\gamma_0)\lambda$. It is easy to see that  $\lambda = \lambda(u_0)$ and hence $\lambda'(u_0) = -\alpha_0 \lambda(u_0)$, $0 = \gamma_0 \lambda$. If $\lambda \neq 0$ then $\alpha_0 = -(\log \lambda(u_0))'$. But this contradicts the assumption  (\ref{eq11}) requesting that $ \frac {\partial \alpha_0 (u_ {1}, u_0, u_ {-1})} {\partial u_ {\pm 1}} \neq 0 $. Hence we have $\lambda =0$. Now consider the equations (\ref{S4_eq14}):
\begin{eqnarray}
&&D_x(\mu) = (\alpha_{1} u_{1,x} + \gamma_{1}) \mu + \beta_{1} u_{1,x} + \delta_{1}, \label{S4_eq15}\\
&&D_x(\tilde{\mu}) = (\alpha_{-1} u_{-1,x} + \gamma_{-1}) \tilde{\mu} - \beta_{-1} u_{-1,x} - \delta_{-1}. \label{S4_eq16}
\end{eqnarray}
From (\ref{S4_eq15}) we obtain that $\mu$ depends only on $u_{1}$ and from (\ref{S4_eq16}) we obtain that $\tilde{\mu}$ depends only on  $u_{-1}$. Now the equations (\ref{S4_eq15}) and (\ref{S4_eq16}) are reduced to the following system of equations:
\begin{eqnarray}  
&&\mu'(u_{1}) = \alpha_{1} \mu(u_1) + \beta_{1}, \quad 0 = \gamma_{1} \mu(u_1) + \delta_{1}, \label{S4_eq17}\\
&&\tilde{\mu}(u_{-1}) = \alpha_{-1} \tilde{\mu}(u_{-1}) - \beta_{-1}, \quad 0 = \gamma_{-1} \tilde{\mu}(u_{-1}) - \delta_{-1}. \label{S4_eq18}
\end{eqnarray} 
By shifting the argument $n$ backwards and forwards by one in the equation (\ref{S4_eq17}) and, respectively, in (\ref{S4_eq18}) we obtain.
\begin{eqnarray}  
&&\mu'(u_{0}) = \alpha_{0} \mu(u_0) + \beta_{0}, \quad 0 = \gamma_{0} \mu(u_0) + \delta_{0}, \label{S4_eq19}\\
&&\tilde{\mu}(u_{0}) = \alpha_{0} \tilde{\mu}(u_{0}) - \beta_{0}, \quad 0 = \gamma_{0} \tilde{\mu}(u_{0}) - \delta_{0}. \label{S4_eq20}
\end{eqnarray}
Now we exlude $ \mu $ and $ \tilde{\mu} $ from these equations and arrive at the differential equation for the function $ \delta_0 $:
\begin{equation} \label{delta0}
\delta_{0,u_0} = \left( \frac{\gamma_{0,u_0}}{\gamma_0} + \alpha_0 \right)\delta_0 - \beta_0 \gamma_0.
\end{equation} 
Equation (\ref{delta0}) is easily solved
\begin{eqnarray}
\fl \delta_0(u_{-1},u_0,u_1) = \frac{1}{4} \frac{1}{(u_0 - u_{-1})(u_0-u_1)} \times\nonumber\\
\fl \times \Bigl( k_1 ( u^2_0 u_1 - 2  u_0 u_{-1} u_1 +  u^2_0 u_{-1}) + 2 k_2( u^2_0 -  u_{-1} u_1)+  2 k_3(- u_1 + 2 u_0 -   u_{-1}) \Bigr) \times\nonumber\\
\fl \times \Bigl(  \tilde{k}_1(  u_0 u_1 -  u_{-1} u_1+  u_{-1} u_0) +2 \tilde{k}_2 u_0 + 2 \tilde{k}_3+4 F_1 (u_{-1}, u_1)(u_0 - u_{-1})(u_0-u_1)\Bigr).\label{delta_first}
\end{eqnarray}
Here $k_1$, $k_2$, $k_3$ and $\tilde{k}_1$, $\tilde{k}_2$, $\tilde{k}_3$  are constants, which appear in the description of functions (\ref{gamma_p3}),  (\ref{eq3_61}) and $F_1 (u_{-1}, u_1)$ is a function to be found. 

Substituting (\ref{delta_first}) into the second equation of (\ref{S4_eq19}), we obtain that $F_1 (u_{-1}, u_1)=\frac{1}{2} \tilde{k}_1$ and
\begin{equation*}
\mu(u_{0}) = \frac{1}{2} \tilde{k}_1 u^2_{0} +\tilde{k}_2 u_{0} + \tilde{k}_3
\end{equation*}
From the second equation of (\ref{S4_eq20})  we get
\begin{equation*}
\tilde{\mu}(u_{0}) = -\frac{1}{2} \tilde{k}_1 u^2_{0} -\tilde{k}_2 u_{0} - \tilde{k}_3.
\end{equation*}

Further, we collect the coefficients for $Y_1$, $Y_{-1}$ in the equality (\ref{S4_eq13}). Substituting the functions $\alpha$, $\beta$, $\gamma$, $\delta$, $\mu$, $\tilde{\mu}$ found above into the  equations obtained  we get identities that
do not give any additional condition on the unknown functions.
Let us collect the coefficients for $Y_0$ in (\ref{S4_eq13}) and find:
\begin{equation*}
D_x(\eta) = R(a_0) - Y_0(h_0) + \mu Y_1(a_0) - \tilde{\mu} Y_{-1}(a_0).
\end{equation*}
Calculating each term and simplifying the last equation, we obtain
\begin{eqnarray*}
\fl D_x(\eta) = \bigl(-\beta_{0,u_0} + \mu \alpha_{0,u_1} - \tilde{\mu} \alpha_{0,u_{-1}} \bigr) u_{0,x} +  \delta_0 \alpha_0 - \delta_{0,u_0} - \gamma_0 \beta_0 + \mu \gamma_{0,u_1} - \tilde{\mu} \gamma_{0,u_{-1}}.
\end{eqnarray*}
A simple analysis of the last equation shows that $\eta$ can depend only on $u_0$. Therefore, the last equation reduces to a system of two equations:
\begin{eqnarray}
&&\eta'(u_0) = -\beta_{0,u_0} + \mu \alpha_{0,u_1} - \tilde{\mu} \alpha_{0,u_{-1}}, \label{Y0eq1}\\
&&0 =   \delta_0 \alpha_0 - \delta_{0,u_0} - \gamma_0 \beta_0 + \mu \gamma_{0,u_1} - \tilde{\mu} \gamma_{0,u_{-1}}.\label{Y0eq2}
\end{eqnarray}
By direct calculation we obtain that the right hand side of (\ref{Y0eq1}) is identically equal to zero.

Investigating the equation (\ref{Y0eq2}) we derive some  additional relations between the constants $k_i$ 我 $\tilde{k}_i$:
\begin{equation}
k_1 = \frac{\tilde{k}_1}{\tilde{k}_2}k_2, \quad k_3 = \frac{\tilde{k}_3}{\tilde{k}_2}k_2.
\end{equation}
Thus, it is proved that if  the decomposition (\ref{exps_R0})
takes place then it should be of the form
\begin{equation}  \label{S4_eq200}
R_0 = \mu Y_{1,0} + \tilde{\mu} Y_{0,-1}.
\end{equation}

Herewith we completely determine the desired coefficients of the quasilinear chain (\ref{eq1})
\begin{equation}  \label{res_latt}
u_{n,xy} =  \alpha_n u_{n,x} u_{n,y} + \beta_n u_{n,x} + \gamma_n u_{n,y} + \delta_n.
\end{equation}
Summarizing the reasonings above we present the explicit expressions for these coefficients
\begin{eqnarray*}
&&\alpha_n=\alpha(u_{n+1},u_n,u_{n-1}) = \frac{1}{u_n - u_{n-1}} - \frac{1}{u_{n+1}-u_n},\\
&&\beta_n=\beta(u_{n+1},u_n,u_{n-1}) = r'(u_n) - r(u_n)\alpha(u_{n+1},u_n,u_{n-1}),\\
&&\gamma_n=\gamma(u_{n+1},u_n,u_{n-1}) = \varepsilon(r'(u_n) - r(u_n)\alpha(u_{n+1},u_n,u_{n-1})),\\
&&\delta_n=\delta(u_{n+1},u_n,u_{n-1}) = -\varepsilon r(u_n) (r'(u_n)-r(u_n)\alpha(u_{n+1},u_n,u_{n-1})),
\end{eqnarray*}
where $r(u_n) = \frac{c_1}{2} u^2_n + c_2 u_n + c_3$ is a  polynomial of the degree not higher then two with arbitrary coefficients and $c_i = \tilde{k}_i$, $\varepsilon = k_2/ \tilde{k}_2$. The boundary conditions reducing the chain to an integrable system of hyperbolic equations are given in the form
\begin{equation}  \label{bc11}
u_{-1}=\lambda, \quad u_{N+1}=\lambda
\end{equation}
where $\lambda$ is a root of the polynomial $r(\lambda)$ i.e. $r(\lambda)=0$.
In the degenerate case when $r(u_n) = c_3$ the boundary conditions are of the form
\begin{equation}  \label{bc21}
u_{-1}=c_3(\varepsilon x+y)+c_4, \quad u_{N+1}=c_3(\varepsilon x+y)+c_5,
\end{equation}
where $c_4$, $c_5$ are arbitrary constants.

Let us investigate the case $ii)$. Assume that some element $R_n$, $n>0$ of the sequence (\ref{seq3})  is linearly expressed in terms of the previous elements:
\begin{equation} \label{eq44_30}
R_n = \lambda R_{n-1} + \cdots,
\end{equation}
but elements $R_k$, $k<n$ are not expressed linearly in terms of the previous elements $R_j$, $j<k$ and $Y_{1,0}, Y_{0,-1}, Y_{1}, Y_0, Y_{-1}$.
Let us apply the operator ${\rm ad}_{D_x}$  to both sides of (\ref{eq44_30}) and obtain
\begin{equation*}
p_n (\lambda R_{n-1} + \cdots) + q_n R_{n-1} + \cdots = D_x(\lambda) R_{n-1} + \lambda (p_{n-1} R_{n-1}+\cdots).
\end{equation*}
We collect the coefficients for the operator $R_{n-1}$ in the resulting equality and find:
\begin{equation*}
D_x(\lambda) = \lambda (p_n - p_{n-1}) + q_n.
\end{equation*}
We substitute explicit expressions for $p_n,p_{n-1},q_n$ into the last equation and get
\begin{equation*}
D_x(\lambda) = -a_0 \lambda - \frac{n^2+n}{2} Y_0(a_0).
\end{equation*}
We substitute the explicit expression for $a_0$ and evaluate $Y_0(a_0)$. So we obtain
\begin{equation}
D_x(\lambda) = -(\alpha_0 u_{0,x} + \gamma_0) \lambda - \frac{n^2+n}{2} \left( (\alpha_{0,u_0} + \alpha^2_0) u_{0,x}  + \gamma_{0,u_0} + \gamma_0 \alpha_0\right)
\end{equation}
It follows from the last equality that  $\lambda$ depends only on $u_0$. Then the equation reduces to a system of two equations
\begin{eqnarray}
&&\lambda'(u_0) = -\alpha_0 \lambda - \frac{n^2+n}{2}(\alpha_{0,u_0} + \alpha^2_0), \label{eq44_32}\\
&&\gamma_0 \lambda + \frac{n^2+n}{2} (\gamma_{0,u_0} + \gamma_0 \alpha_0) = 0.
\end{eqnarray}
Rewrite the equation (\ref{eq44_32}) as follows
\begin{equation}
\lambda'(u_0) = \frac{-\lambda(u_0)(2u_0-u_1 -u_{-1})-(n^2+n)}{(u_0-u_{-1})(u_0 - u_1)}
\end{equation}
or
\begin{equation}
\lambda'(u_0)(u^2_0 - u_0 u_1 - u_{-1}u_0+ u_1 u_{-1}) = -\lambda(u_0)(2u_0-u_1 -u_{-1}) - (n^2+n).
\end{equation}
Since the variables $u_{-1},u_0,u_1$ are independent then  the last equation implies immediately that $\lambda = 0$ and $n^2+n=0$. Thus we have $n=0$ or $n=-1$. Both solutions contradict the assumption $n>0$. Therefore the case $ii)$ is never realized.

Up to point trnsformations there are three essentially different versions of the chain (\ref{res_latt}):


1) If $c_1=c_2=0$, then by the shift transformation $u\rightarrow u-c_3(\varepsilon x+y)$ the chain (\ref{res_latt}) reduces to the known 
Ferapontov-Shabat-Yamilov  chain (see \cite{Fer-TMF, ShY})
\begin{equation}  \label{var1}
u_{n,xy}=\alpha_nu_{n,x}u_{n,y},
\end{equation}

2) If $c_1=0$, $c_2\neq0$, then by shifting $u \rightarrow u- \frac{c_3}{c_2}$ and stretching $x \rightarrow \frac{x}{\varepsilon c_2}$, $y \rightarrow \frac{y}{c_2}$ we obtain the chain
\begin{equation}  \label{var2}
u_{n,xy}=\alpha_n(u_{n,x}u_{n,y}-u_n(u_{n,x}+u_{n,y})+u_n^2) +u_{n,x}+ u_{n,y} -u_n,
\end{equation}

3) For  $c_1\neq0$ by the shift transformation $u \rightarrow u - \frac{c_2}{c_1}$ and by the stretching $x \rightarrow \frac{2}{\varepsilon c_1}x$, $y \rightarrow \frac{2}{c_1} y$ chain (\ref{res_latt})  can be reduced to the form
\begin{equation}  \label{var3}
u_{n,xy}=\alpha_n(u_{n,x}u_{n,y}-s_{n}(u_{n,x}+u_{n,y})+s_{n}^2) + s'_{n}(u_{n,x}+u_{n,y}-s_n),
\end{equation}
where $ s_{n}=u_n^2+C$ and $C=\frac{c_3}{c_1}-\left(\frac{c_2}{c_1}\right)^2$ -- is an arbitrary constant.

Thus we have proved that any chain integrble in the sense of our Definition 1 is of the form (\ref{res_latt}). In order to complete the proof of Theorem~\ref{th0} we have to verify the converse statement. It is done in the following theorem.

\begin{theorem} \label{MainTheorem}
The chain (\ref{res_latt}),  found as a result of the classification, is integrable in the sense of Definition 1 formulated in the Introduction.
 \end{theorem}

We introduce special notations for multiple commutators of the operators $\left\{Y_i\right\}$
\begin{equation}\label{multi}
Y_{i_k,\dots ,i_0}=[Y_{i_k},Y_{i_{k-1},\dots ,i_0}].
\end{equation}
The structure of the Lie algebra generated by the operators $\left\{Y_i\right\}$ can be studied by the method developed in our previous paper \cite{HabPoptsovaSIGMA17}. One can prove that any element of this algebra can be represented as a linear combination of the following operators
\begin{equation} \label{basis0}
Y_i,\,Y_{i+1,i},\,Y_{i+2,i+1,i},\dots .
\end{equation}
It follows from the formula (\ref{eq2_5}) that the algebra $\mathcal{L}(y,N)$ corresponding to the system  (\ref{eq2_1}) is an extension of this algebra, obtained by adding one more generator, namely, the operator $R$.

Recall that in the paper \cite{HabPoptsovaSIGMA17} the particular case of a chain (\ref{res_latt}) was studied in detail. Namely, the following theorem was proved:
 
 \begin{theorem} \label{Theorem4}
The chain
\begin{equation} \label{res_latt0}
 u_{n,xy} =  \left( \frac{1}{u_n - u_{n-1}} - \frac{1}{u_{n+1}-u_n} \right)u_{n,x} u_{n,y}
 \end{equation}
is integrable by Definition 1, formulated in the Introduction.
\end{theorem} 
Recall briefly the scheme of the proof of the Theorem~\ref{Theorem4}. The basis
\begin{equation} \label{basis}
 \{Y_i\}_{i=0}^{N},\quad \{Y_{i+1,i}\}_{i=0}^{N-1}, \quad \{Y_{i+2,i+1,i}\}_{i=0}^{N-2},\quad \ldots, Y_{N,N-1,\ldots,0}.
\end{equation}
was constructed on the set of multiple commutators of the operators $Y_0,...,Y_N$ corresponding to the chain (\ref{res_latt0}).

In order to prove that there is a basis (\ref{basis}) on the set of multiple commutators of the operators $Y_0,...,Y_N$ corresponding to the chain (\ref{res_latt}) we can repeat the proof of the Theorem~5.2 from the paper \cite{HabPoptsovaSIGMA17} (see Appendix), putting $a_i = \alpha_i u_{i,x}+\gamma_i$. These proof is cumbersome, so  we do not give it here.

In order to prove the main Theorem~\ref{MainTheorem} we consider the algebra Lie $\mathcal{L}(y,N)$ generated by the operators $Y_0,...,Y_N, R$ and we prove
that the finite basis exists in this algebra
\begin{equation} \label{I766}
R, \quad \{Y_i\}_{i=0}^{N},\quad \{Y_{i+1,i}\}_{i=0}^{N-1}, \quad \{Y_{i+2,i+1,i}\}_{i=0}^{N-2},\quad \ldots, Y_{N,N-1,\ldots,0}.
\end{equation}
So it remains to verify that any multiple commutator of the operator $R$  with the operators (\ref{basis}) is linearly expressed in terms of the operators from the set (\ref{I766}).
 
Let us prove the Theorem~\ref{MainTheorem}. 
 
Proof. Here we consider truncated chains, i.e. finite systems of the hyperbolic equations (\ref{eq2_1}), obtained 
by imposing cut-off conditions to the initial chain.  Note that commutation relations near the cut-off points change in the transition from an infinite chain to a truncated one.

We prove the Theorem~\ref{MainTheorem} by induction. Let us justify the induction base. 
The first step of the proof requires the following formulas: 
\begin{eqnarray}
&&\left[ D_x, \bar{R}_0 \right] = - a_0 \bar{R}_0 + h_1 Y_{1,0} - Y_0(h_1) Y_1 + \bigl( R(a_0) - Y_0(h_0) \bigr)Y_0, \label{eq4_30}\\
&&\left[ D_x, \bar{R}_N \right] = - a_N \bar{R}_N - h_{N-1}\left[Y_N, Y_{N-1} \right] -\nonumber\\
&&\hphantom{\left[ D_x, \bar{R}_N \right] =}  - Y_N(h_{N-1})Y_{N-1} + \bigl( R(a_N) - Y_N(h_N) \bigr)Y_N, \label{eq4_31}\\
&&\left[D_x, \bar{R}_k \right] = - a_k \bar{R}_k - h_{k-1} \left[ Y_k, Y_{k-1} \right] + h_{k+1} \left[ Y_{k+1}, Y_k \right] -\nonumber\\
&&\hphantom{\left[D_x, \bar{R}_k \right] =}  - Y_k(h_{k-1})Y_{k-1} + \bigl( R(a_k) - Y_k(h_k) \bigr)Y_k - Y_k (h_{k+1})Y_{k+1}. \label{eq4_32}
\end{eqnarray}
Here $\bar{R}_j=[Y_j,R]$, $j=0,1\dots ,N$. 

At first we study the end points $k=0$ and $k=N$. Let us show that the following equality holds
\begin{equation} \label{eq4_33}
\bar{R}_0 = \lambda^{(0)}R + \mu^{(0)}Y_{1,0} + \nu^{(0)} Y_1 + \eta^{(0)} Y_0.
\end{equation}
We apply the operator $ {\rm ad} _ {D_x} $ to both sides of the equality (\ref{eq4_33}) and simplify using (\ref{DxY3}), (\ref{DxY10}), (\ref{DxR}), (\ref{eq4_30}), as a result we obtain
\begin{eqnarray} 
\fl -a_0(\lambda^{(0)} R +\mu^{(0)} Y_{1,0}+\cdots) + h_1 Y_{1,0} +\cdots=\nonumber\\
= D_x(\lambda^{(0)})R + D_x(\mu^{(0)})Y_{1,0} + \mu^{(0)}(-(a_1+a_0)Y_{1,0}+\cdots). \label{eq4_34}
\end{eqnarray}
Here three dots stand for a linear combination of the operators $Y_1$, $Y_0$. Collecting the coefficients for the operators $R$ and $Y_{1,0}$ in (\ref{eq4_34}), we obtain a system of the equations
\begin{eqnarray}
&&D_x(\lambda^{(0)}) = -a_0 \lambda^{(0)}, \label{eq4_35}\\
&&D_x(\mu^{(0)}) = a_1 \mu^{(0)} + h_1. \label{eq4_36}
 \end{eqnarray} 
The equation (\ref{eq4_35}) coincides with equation (\ref{S4_eq13}) for $i=0$, hence $\lambda^{(0)} = 0$. The equation (\ref{eq4_36}) coincides with the first equation (\ref{S4_eq14}) for $i=0$, hence $\mu^{(0)} = \mu$. It is easy to show that $\nu^{(0)} = \eta^{(0)} =0$. Thus we have proved that the decomposition (\ref{eq4_33}) has the form
\begin{equation} \label{R0}
\bar{R}_0 = \mu^{(0)} Y_{1,0}.
\end{equation}

Let us show that the following equality holds
\begin{equation}  \label{eq4_37}
\bar{R}_N = \lambda^{(N)}R + \tilde{\mu}^{(N)}Y_{N,N-1} + \eta^{(N)} Y_N + \epsilon^{(N)} Y_{N-1}.
\end{equation}
We apply ${\rm ad}_{D_x}$ to both sides of the relation (\ref{eq4_37}):
\begin{eqnarray}
\fl -a_N(\lambda^{(N)} R +\tilde{\mu}^{(N)} Y_{N,N-1}+\cdots) - h_{N-1} Y_{N,N-1} + \cdots =\nonumber\\
= D_x(\lambda^{(N)})R + D_x(\tilde{\mu}^{(N)})Y_{N,N-1} + \tilde{\mu}^{(N)}(-(a_N+a_{N-1})Y_{N,N-1}+\cdots). \label{eq4_38}
\end{eqnarray}
Here three dots stand for a linear combination of the operators $Y_N$, $Y_{N-1}$. Collecting the coefficients for $R$ and $Y_{N,N-1}$, we get the system:
\begin{eqnarray}
&&D_x(\lambda^{(N)}) = -a_N \lambda^{(N)}, \label{eq4_39}\\
&&D_x(\tilde{\mu}^{(N)}) = a_{N-1} \tilde{\mu}^{(N)} - h_{N-1}. \label{eq4_40}
\end{eqnarray}
The equation (\ref{eq4_39}) coincides with equation  (\ref{S4_eq13}) for $i=N$, hence $\lambda^{(N)} = 0$. The equation (\ref{eq4_40})  coincides with the second equation (\ref{S4_eq14}) for $i=N$, hence we have $\tilde{\mu}^{(N)} = D^N_n \tilde{\mu}(u_{-1}) = \tilde{\mu}(u_{N-1})$. It is easy to show that $ \eta^{(N)} = \epsilon^{(N)}=0$. So we have proved that the decomposition (\ref{eq4_37}) has the form:
\begin{equation}   \label{RN}
\bar{R}_N =  \tilde{\mu}^{(N)}Y_{N,N-1}.
\end{equation}

Now we concentrate on the inner points by taking $k$ from the set $0<k<N$. Let us show that the following equality holds
\begin{equation} \label{eq4_41}
\bar{R}_k = \lambda^{(k)} R + \mu^{(k)} Y_{k+1,k} +\tilde{\mu}^{(k)} Y_{k,k-1} + \nu^{(k)} Y_{k+1}+\eta^{(k)} Y_{k} + \epsilon^{(k)} Y_{k-1}.
\end{equation}
We apply the operator ${\rm ad}_{D_x}$ to both sides of the relation (\ref{eq4_41}):
\begin{eqnarray} 
\fl -a_k( \lambda^{(k)} R + \mu^{(k)} Y_{k+1,k} +\tilde{\mu}^{(k)} Y_{k,k-1}+\cdots) - h_{k-1} Y_{k,k-1} + h_{k+1} Y_{k+1,k} + \cdots= \nonumber\\
= D_x(\lambda^{(k)})R +D_x(\mu^{(k)})Y_{k+1,k}+ D_x(\tilde{\mu}^{(k)})Y_{k,k-1} +\nonumber\\
+ \mu^{(k)}(-(a_{k+1}+a_{k})Y_{k+1,k}+\cdots) +\tilde{\mu}^{(k)}(-(a_k+a_{k-1})Y_{k,k-1}+\cdots). \label{eq4_42}
\end{eqnarray}
Here three dots stand for a linear combination of the operators $Y_0, Y_1, ..., Y_{N-1}, Y_N$.
Collecting the coefficients for $R$, $ Y_{k+1,k}$, $Y_{k,k-1}$ in  (\ref{eq4_42}), we obtain the system
\begin{eqnarray}
&&D_x(\lambda^{(k)}) = -a_k \lambda^{(k)}, \label{eq4_43}\\
&&D_x(\mu^{(k)}) = a_{k+1}\mu^{(k)} + h_{k+1}, \label{eq4_44}\\
&&D_x(\tilde{\mu}^{(k)}) =  a_{k-1} \tilde{\mu}^{(k)}- h_{k-1}. \label{eq4_45}
\end{eqnarray}
The equation (\ref{eq4_43}) coincides with (\ref{S4_eq13}) if $i=k$. That is why we obtain that $\lambda^{(k)} = 0$. The equation (\ref{eq4_44}) coincides with the first equation   (\ref{S4_eq14}) if $i=k$, and equation (\ref{eq4_45}) coincides with the second equation (\ref{S4_eq14}) if $i=k$. Hence, $\mu^{(k)} = D^k_n(\mu(u_1)) = \mu(u_{k+1})$, $\tilde{\mu}^{(k)} = D^k_n(\tilde{\mu}(u_{-1})) = \tilde{\mu}(u_{k-1})$. It is easy to show that $\nu^{(k)}=\eta^{(k)}=\epsilon^{(k)}=0$. Thus, we have proved that the decomposition (\ref{eq4_41}) has the form:
\begin{equation}  \label{Rk}
\bar{R}_k =\mu^{(k)} Y_{k+1,k} +\tilde{\mu}^{(k)} Y_{k,k-1}.
\end{equation}

Now we calculate the commutator $\left[Y_{i+1,i}, R\right]$ for some $i$,  $0 \leq i \leq N-1$. Using the Jakobi identity, we obtain
\begin{eqnarray*}
&&\left[ Y_{i+1,i}, R  \right] = - \left[ R, Y_{i+1,i}  \right] = -\left[ R, \left[ Y_{i+1}, Y_i \right]\right] = \\
&&= \left[Y_{i+1}, \left[ Y_i, R \right] \right] + \left[Y_i, \left[ R, Y_{i+1} \right] \right] =\\
&&= \left[Y_{i+1}, \mu^{(i)} Y_{i+1,i} + \tilde{\mu}^{(i)} Y_{i,i-1} \right] - \left[ Y_i, \mu^{(i+1)} Y_{i+2,i+1} + \tilde{\mu}^{(i+1)} Y_{i+1,i} \right] =\\
&&=\Lambda^{(i)} Y_{i+2,i+1,i} + M^{(i)} Y_{i+1,i,i-1} + \kappa^{(i)}Y_{i+2,i+1} + \eta^{(i)}Y_{i+1,i} +\zeta^{(i)}Y_{i,i-1},
\end{eqnarray*}
where $\Lambda^{(i)}, M^{(i)}, \kappa^{(i)}, \eta^{(i)},\zeta^{(i)}$ -- some functions that depend on dynamic variables. Herewith $\zeta^{(0)} =0$, $M^{(0)} = 0$, $\Lambda^{N-1} = 0$, $\kappa^{N-1} = 0$.

Now let us justify the inductive transition. Assume that for a given  $M$, $0 \leq k<M \leq N-1$ the following formula holds:
\begin{eqnarray}
&&\left[ Y_{M,M-1,...,k}, R \right]=\Lambda Y_{M+1,M,M-1,...,k} + M Y_{M,M-1,...,k,k-1} + \nonumber\\
&&+\nu Y_{M,M-1,...,k} + \varepsilon Y_{M+1,M,M-1,..,k+1} +\eta Y_{M-1,...,k,k-1} +\nonumber\\
&&+ \zeta Y_{M-1,M-2,...,k} + \theta Y_{M,M-1,...,k+1} + \xi Y_{M-2,M-2,...,k-1} + \cdots +\nonumber\\
&&+ \cdots + \kappa Y_{M+1,M} + \varphi Y_{M,M-1} + \cdots + \chi Y_{k,k-1}. \label{IndStep}
\end{eqnarray}
Let us show that a similar representation holds for $M+1$.
Using the Jakobi identity, we obtain that the following decomposition holds:
\begin{eqnarray*}
&&\left[ Y_{M+1,M,M-1,...,k}, R  \right] = -\left[ R, \left[Y_{M+1}, Y_{M,M-1,...,k}\right]\right] =\\
&&=\left[Y_{M+1}, \left[Y_{M,M-1,...,k},R\right]\right] +   \left[ Y_{M,M-1,..,k}, \left[R,Y_{M+1}\right]\right] = \\
&&=\left[Y_{M+1}, \left[Y_{M,M-1,...,k},R\right]\right] -  \left[ Y_{M,M-1,..,k}, R_{M+1}\right].
\end{eqnarray*}
We substitute the decomposition (\ref{IndStep})  and a proper one of the equations (\ref{Rk}), (\ref{R0}) or (\ref{RN}) (it depends on the concrete value of $M$: $M=0$, $M=N$ or $0<M<N$) into the last formula. Then we expand the commutators using the linearity property. The latter completes the proof of Theorem~\ref{MainTheorem}.

\section*{Conclusions}

In this paper the problem of the integrable classification of two-dimensional chains of the type (\ref{eq0}) is studied. For chains of a special type (\ref{eq1}), (\ref{eq11}) a complete description of the integrable cases is obtained. By integrability of the  chain we mean here the existence  of reductions in the form of arbitrarily high order systems of hyperbolic type equations  that are Darboux integrable.
In the list obtained, along with the known ones, there are new chains (see Chains ii) and iii) in Theorem 1).

The algorithm used for classification is relatively new, testing this algorithm is one of the goals of the work. It is based on the concept of the characteristic Lie algebra applied earlier to the systems of hyperbolic type equations with two independent variables (see, for instance, \cite{ZMHSbook}, \cite{ZMHS-UMJ} and the references therein). It is well known that the characteristic algebras in both directions for a Darboux integrable system have finite dimension. In the present article we adopted the concept to the classification of 1+2 -dimensional lattices.

As the examples show (see \cite {ZhiberMurtazina} - \cite {Million}), the characteristic algebras of the hyperbolic systems with two independent variables integrable by means of the inverse scattering method are slow growth algebras.

\section*{Acknowledgments}

The authors gratefully acknowledge financial support from a Russian Science Foundation
grant (project 15-11-20007).

\end{document}